# Observation of non-classical light in semiconductor microcavities

M.V. Lebedev*, A.L. Parakhonsky and A.A. Demenev

*Institute of Solid State Physics, Russian Academy of Sciences,
142432, Chernogolovka, Moscow distr., Russia*

Semiconductor microcavities are widely used to study collective interactions of cavity exciton-polaritons leading to their condensation phenomenon. Exciton – light interaction is highly enhanced in such structures due to the resonance enhancement of the electromagnetic field inside the high quality cavity. Considerably lower interest is concerned to exciton states non-resonant with the cavity mode for which the exciton – photon interaction is strongly reduced due to the existence of the cavity. We point out here that these states are responsible for non-classical light generation in semiconductor microcavities.

**Introduction**

Non-classical states of light are of great interest because of their potential applications in modern optics, quantum cryptography, high precision optical measurements, quantum informatics and telecommunication. In addition those states excite also fundamental interest for better understanding of the nature of light. Some possible definitions of what is called non-classical light can be given. The one most commonly used is that non-classical state of light cannot be described with a positively defined Glauber-Sudarshan P-representation function. The second order correlation function $g^{(2)}$ for a non-classical light field can be less than unity what is forbidden for light from a usual thermal light source. Such a non-classical light state observed experimentally was fluorescence light from a single atom demonstrating $g^{(2)}(0)=0$ called antibunching phenomenon [1]. A lot of beautiful experiments were made with two-photon correlated light states called also biphotons [2], which are photon pairs generated in nonlinear optical crystals by a spontaneous down conversion process. Squeezed states of light discovered in 1987 [3] demonstrate phase sensitive photon noise which can be produced as sub-Poissonian. This is promising for increasing accuracy of optical measurements depending on photon noise. The existing sources of non-classical light are rather weak. Single atom sources give something like $10^6$ cps, a nonlinear optical crystal under pumping with a 1W laser beam gives $10^3$-$10^4$ pairs per second and a specially constructed two-photon light source consisting of two coupled resonators with quantum dots inside gives up to $10^7$ cps [4]. We describe here a source of non-classical light based on a semiconductor microcavity, which gives 0.3 mW of nonclassical light; this corresponds to $1.3\times10^{15}$ photons per second, under laser pumping of 10 mW.

*Corresponding author. Tel./Fax: +7 496 5228231/+7 496 5228160.
E-mail address: lebedev@issp.ac.ru (M.V. Lebedev).

**Experimental**

The microcavity (MC) structure grown by a metalorganic vapor-phase epitaxy technique has top (bottom) Bragg reflectors composed of 17 (20) $\lambda/4$ $Al_{0.13}Ga_{0.87}As$/AlAs layers. The $3\lambda/2$ GaAs cavity contains six 10 nm-thick $In_{0.06}Ga_{0.94}As$/GaAs quantum wells. The Rabi splitting of the structures is $\Omega \approx 6$ meV. A gradual variation of an active layer thickness along the sample provides a change in the photon mode energy, so the detuning between the exciton and the cavity mode could be changed simply by changing the position of a laser spot on a crystal surface. The cavity Q-factor was in the order of 3000. Experiments have been carried out on several regions of the sample with negative detuning while full change of detuning on length of the sample was in the range from -6 to 6 meV.

Our experimental setup is shown in Fig.1. A GaAs MC sample was mounted on a cold finger of an optical helium cryostat. It was excited with a semiconductor laser which generated a rather broad spectrum (2 nm) of several transverse modes. Laser radiation was focused by a lens with focal length of 45 mm onto a crystal surface into a spot with angle divergence of 17 mrad. The distance between the lens and the sample was 610 mm. Laser radiation was well polarized in a horizontal plane (an incidence plane) and it was falls on a crystal surface at an angle of $4^0$ from the normal. A diaphragm $D_1$ with a diameter of 3 mm was positioned just before the crystal to reduce strayed laser light. Laser diode was mounted on a Peltier cooler and its generation wavelength could be tuned by the cooling. Two multimode fibers coupled to lenses were mounted on the arms of a home-made goniometer, which enabled us to tune and control the direction of observation of the transmitted light and the one scattered through the microcavity. Two diaphragms $D_2$ and $D_3$ with diameters of 5 mm were positioned in front of the lenses to achieve good angle resolution of the registered light. The distances between the sample and the diaphragms were 525 mm for the arms of our goniometer. The fibers were connected to two avalanche photodiodes with good single photon response. Correlation measurements were carried out with a standard start-stop setup including two discriminators, time-to-amplitude converter and a multichannel analyzer.

A broad band light source[1] was focused with a microscopic objective on a crystal surface into a spot of 100 mkm in diameter by means of the mirror M2 at an incidence angle of 10 from the normal. This gave us a possibility of express-testing the local detuning between the exciton and cavity mode at different points of our sample, simply connecting our fibers to two USB-spectrometers (not shown in Fig.1) instead of avalanche photodiodes. The spectra of radiation detected with avalanche photodiodes were controlled in the same way.

**Results and discussion**

We revealed that the second order correlation function $g^{(2)}$ demonstrate surprisingly long oscillations between correlated and anti-correlated photon states (see Fig.2). To avoid any doubt about the possible source of correlations in our electronics we changed the length of the optical fiber in the stop channel of our setup. This led to the appropriate shift of the zero position $\tau_0$ of our correlation function (approximately by 20 ns). We expected pronounced deviations of $g^{(2)}$ from unity only for the case $k_1 = -k_2$, where $k_1 = \frac{\varepsilon_1}{\hbar c} \cdot \sin(\alpha_1)$ and $k_2 = \frac{\varepsilon_2}{\hbar c} \cdot \sin(\alpha_2)$ are the wave vector components of the detected photons parallel to the crystal plane. However, we found that this condition is not necessary for observing strong correlations. The spectra of correlated photons also need not coincide exactly. All the scattered radiation belonging to different parts of the low energy polariton branch demonstrates similar correlations. The detailed analysis of correlation dependence on angles of scattering as well as spectra of detected photons will be done in the following studies. The only thing seems to follow from our measurements up to now. I. e. the best visibility of the correlation function can be observed just when the condition $k_1 = -k_2$ is fulfilled, the spectra of detected photons coincide exactly and the observed polariton states are not too close to the exciton energy but lie on a photon like part of the lower polariton branch far enough from the $k=0$ state to avoid registration of direct laser light.

---
[1] A light emitting diode (LED) with the characteristic spectral width equal to 20nm.

Second order correlation function depends on the laser pumping power. The period of oscillations definitely increases with the decrease of the pumping but this dependence is different at different points of the crystal surface. At some points we observed a strong dependence on the pumping power: the period of oscillations increased up to three times with the pumping reduction from 10 to 0.3 mW. But in most cases the increase of the period was not so pronounced and was about 10%. It is more or less clear that this period may depend both on the local cavity quality and detuning. Additional detailed experiments should be done to clarify this question.

The observed oscillations disappeared as a result of the excitation of the microcavity with a chaotic light source (see Fig.3). As a chaotic light source LED with a broad spectrum with characteristic width of 20 nm was used. We got one more unexpected result while measuring the autocorrelation function of the light transmitted through our MC sample. This was done using only one arm of our goniometer positioned directly into the transmitted laser beam. The input fiber was connected to an X-type fiber splitter, the outputs of which were connected to our detectors. We measured thus the $g^{(2)}(k_1 = k_2 = k_{laser}, \tau)$ function. We expected to observe the absence of any deviations from unity which characterizes the Poissonian photon statistics of laser light. This experimental result is shown in Fig.4. Pronounced correlations remain also in the transmitted light beam, though the visibility of oscillations: $\mathbf{V} = \frac{g^{(2)}_{max} - g^{(2)}_{min}}{g^{(2)}_{max} + g^{(2)}_{min}}$ is not as high as in Fig.2.

Measurements of the temperature dependence of $g^{(2)}$ showed the role of the cavity. It is well known that exciton position exhibits red shift with increasing temperature. We started with the situation of negative detuning at our lowest working temperature and observed the visibility of $g^{(2)}$ with the increase of temperature while the exciton made a move towards the cavity mode position. The result is shown in Fig.5. The visibility tends abruptly to zero at the point of energy crossing of exciton and the cavity mode, i.e. at zero detuning. This fact has a simple explanation. At zero detuning, extended exciton states can be excited with the cavity field. This states exhibit strong relaxation and as a result strong absorption of light by the cavity sets in. In fact cavity influence in this case vanishes because multiple beam reflection from the mirrors is not possible due to high absorption.

The second order correlation function of light emitted by a semiconductor microcavity was calculated in [5]. The cavity mode in this work is supposed to be exactly in resonance with excitons and any non-resonant interactions are neglected. This assumption is often made when considering intracavity electrodynamics, but it is obviously not our case. Nevertheless the result of calculations looks in general very similar to experimentally observed one. Just as we observe it exhibits Rabi oscillations between bunched and antibunched photon states, but cannot explain the long times of these oscillations. We can interpret our findings as Rabi oscillations between the ground state of the crystal and states of localized excitons; their interaction with the intracavity field is strongly reduced because they can interact only with non-resonant field states of the cavity [6]. These exciton states have the interaction with non-resonant electromagnetic field in the cavity as the only channel of relaxation. The weakness of this interaction explains the long times of the Rabi oscillations observed in our experiments. The polariton modes which we observe in our experiments are coupled to the non-resonant intracavity field states by the weak process of Rayleigh scattering of the intracavity light. This explanation deserves of course further verifications which are in progress but it seems to give a good basis for new experiments. If further experiments confirm our explanation it will mean that off-resonant localized exciton states can accumulate a considerable amount of energy of a pumping laser and play a very important role in understanding of light-exciton interaction in semiconductor microcavities.

Very long oscillation times observed in our experiments bring forth thus that the correlation observed should be attributed not to the light field only but to the whole light-exciton system. The propagation time between our sample and detectors is considerably less than the correlation times observed for the Rabi oscillations. In simple words, we observe correlations between two photons one of which is just detected and the other is not emitted at that moment. A natural question thus arises: does the detection of the first photon affect the emission of the second? It seems that the answer could be given after detailed quantum-mechanical analysis of the procedure of measurements on the basis of some model including non-resonant intracavity light states and localized excitons.

Our results can be considered also from a practical point of view. The intensity of the transmitted light can be easily measured with a power meter. Our measurements showed that output power depends linearly on the input power of laser radiation. We obtained the output power of 0.3 mW under the laser pumping power of 10 mW. This means that we have about 0.3 mW of non-classical light which corresponds to $1.3 \times 10^{15}$ photons per second. So, we can conclude that one can use semiconductor microcavities as a very bright source of non-classical light.

**Captions**

Fig.1. The experimental setup. $M_1$, $M_2$ – mirrors, $D_1$, $D_2$, $D_3$ – diaphragms, $f_1$, $f_2$ – fibers, $APD_1$, $APD_2$ – avalanche photodiodes, MCA – multichannel analyzer, TAC – time-to-amplitude converter.

Fig.2. The normalized intensity correlation function $g^{(2)}(k_1,k_2,\tau)$. $\alpha_1 = 13.5°$, $\alpha_2 = 11°$, $T = 12K$. The visibility $\mathbf{V} = \frac{g^{(2)}_{max} - g^{(2)}_{min}}{g^{(2)}_{max} + g^{(2)}_{min}} \approx 0.82$. Excitation power $P = 10$ mW. The dashed curve shows the empirical approximation formula: $g^{(2)}(\tau) = 1 + Ae^{-\frac{|\tau-\tau_0|}{t}} \cos\frac{2\pi(\tau-\tau_0)}{t_{Rabi}}$ with the following parameters: $A = 1.257$, $t = 270.27$ ns, $\tau_0 = 67.33$ ns, $t_{Rabi} = 61.05$ ns. Strong bunching at zero delay: $g^{(2)}(\tau = \tau_0) = 2.55$ and long lived Rabi oscillations with the period of 61 ns are evident. Cavity detuning was δ = -2 meV.

Fig.3. The normalized intensity correlation function $g^{(2)}$ under excitation of the sample with a chaotic light source (LED). Any deviations from unity are absent. δ = - 3 meV. $\alpha_1 = \alpha_2 = 1°$, $T = 12K$. $P = 0.027$ mW.

Fig.4. The normalized intensity autocorrelation function $g^{(2)}(k_1 = k_2 = k_{laser},\tau)$ under laser excitation. $T = 12K$. The visibility $\mathbf{V} = 0.45$. The excitation power $P = 0.82$ mW. δ = - 2 meV. The dashed curve shows the empirical approximation formula: $g^{(2)}(\tau) = 1 + Ae^{-\frac{|\tau-\tau_0|}{t}} \cos\frac{2\pi(\tau-\tau_0)}{t_{Rabi}}$ with the following parameters: $A = 0.65$, $t = 303$ ns, $\tau_0 = 71.52$ ns, $t_{Rabi} = 60.08$ ns.

Fig.5. Temperature dependence of the visibility of $g^{(2)}(k_1,k_2,\tau)$ function and of the positions of polariton lines. Visibility goes to zero at zero detuning. $\alpha_1 = 10°$, $\alpha_2 = 8°$.

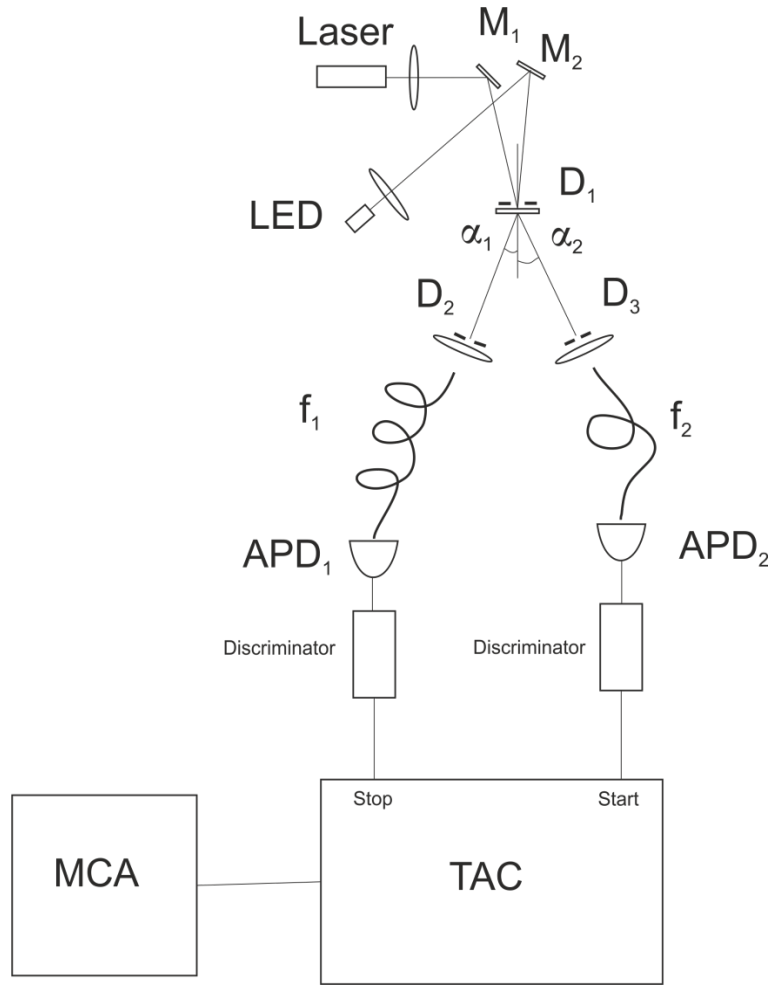

Fig.1

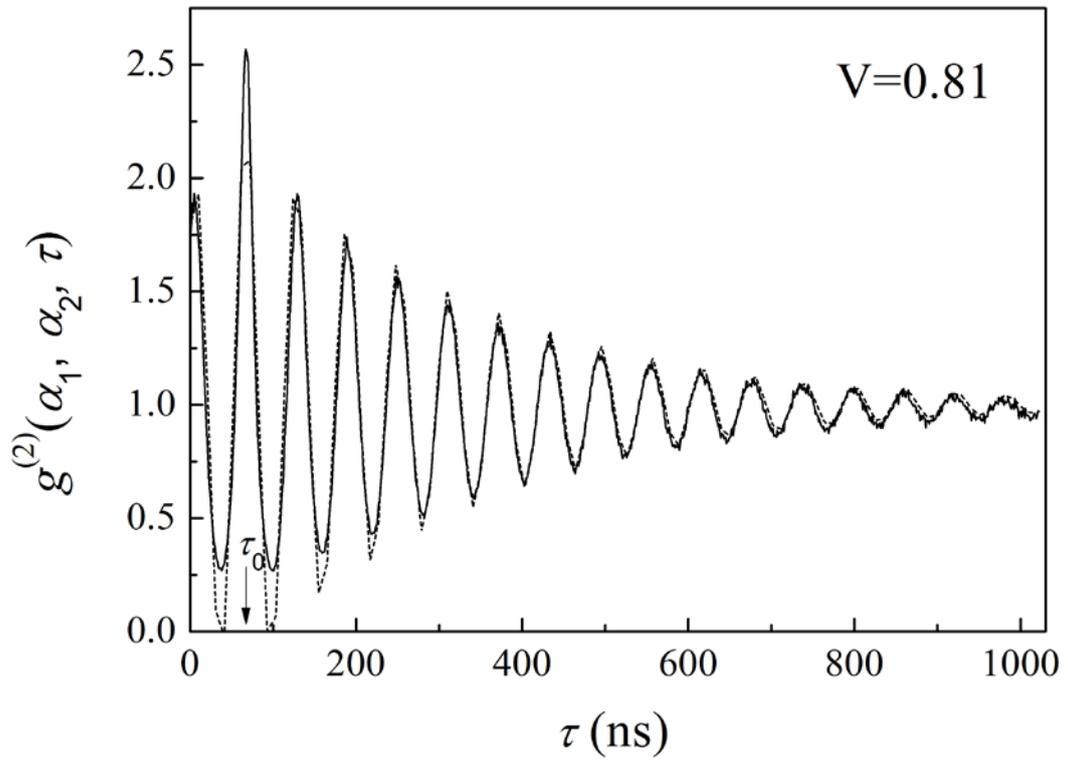

Fig.2

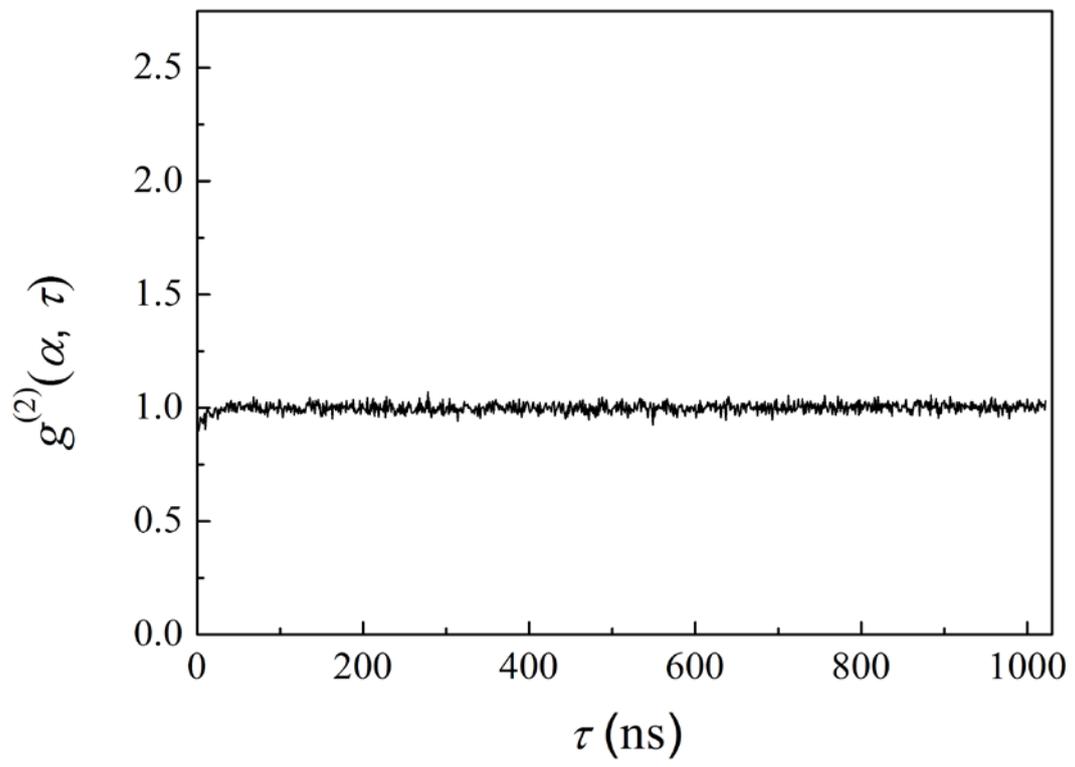

Fig.3

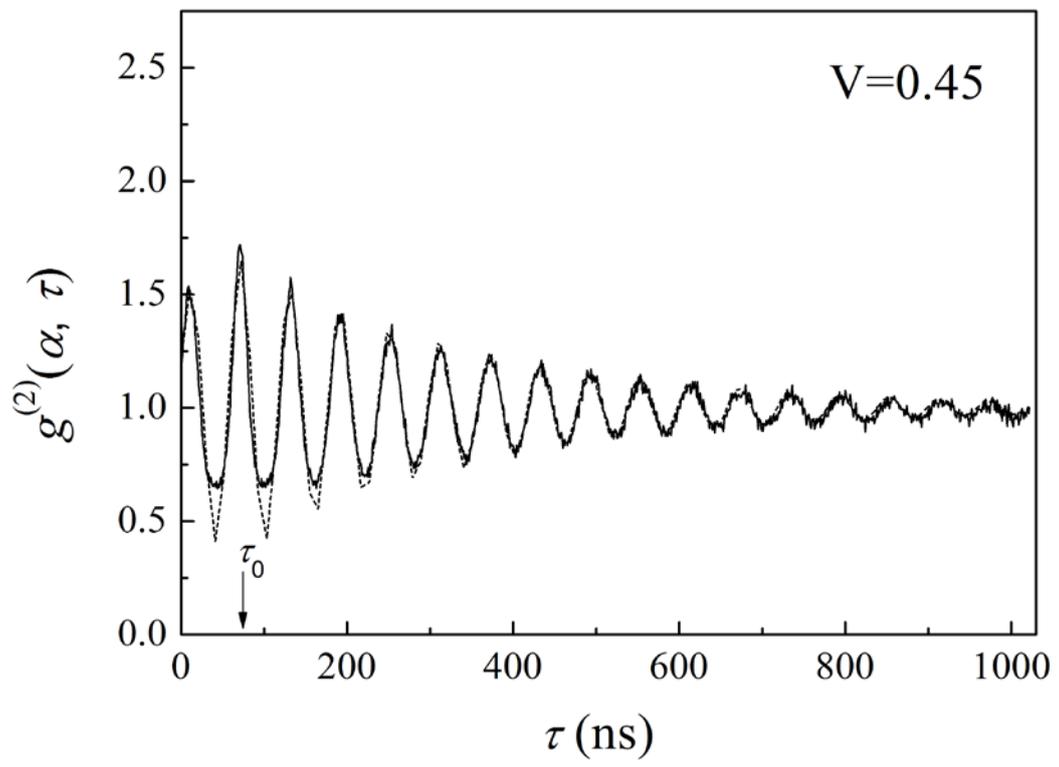

Fig.4

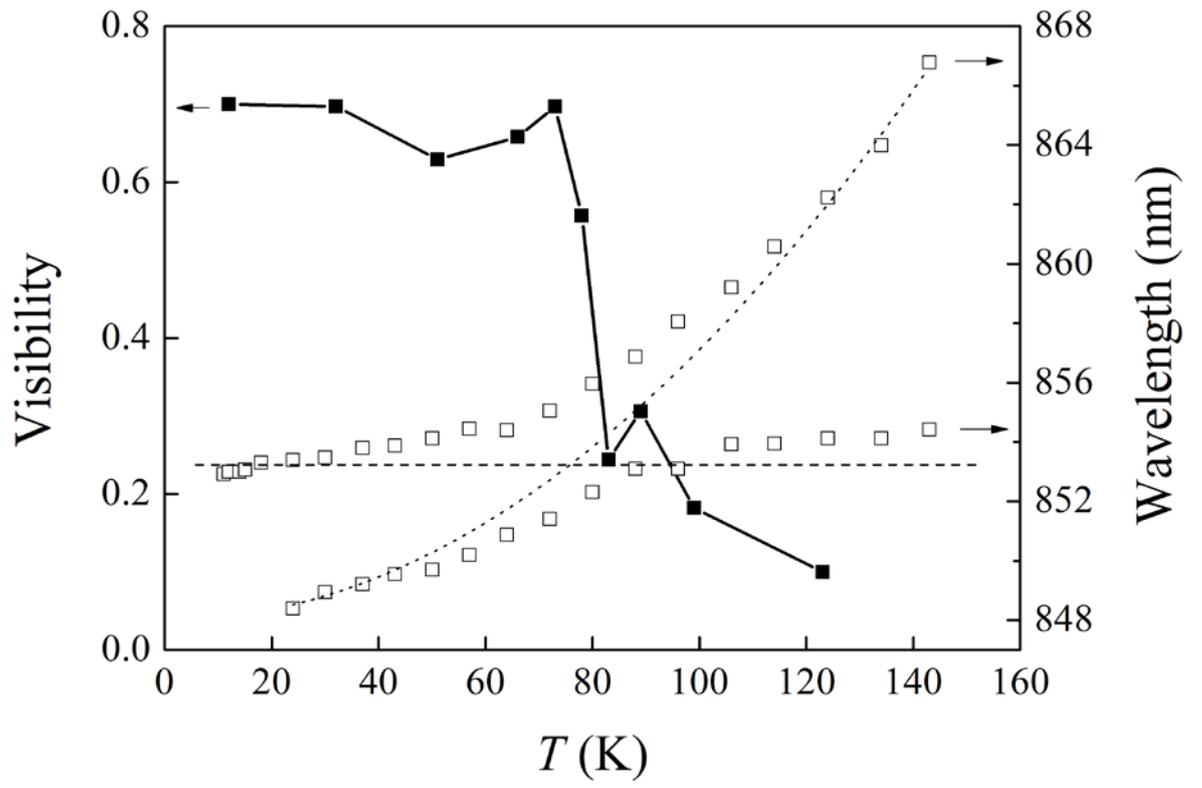

Fig.5